\documentclass[prl, a4paper, amsfonts, amssymb, amsmath, reprint, showkeys, nofootinbib, twoside]{revtex4-1}
\usepackage{lipsum}
\usepackage[english]{babel}
\usepackage[mathcal]{eucal}
\usepackage[utf8]{inputenc}
\usepackage[colorinlistoftodos, color=green!40, prependcaption]{todonotes}
\usepackage[pdftex, pdftitle={Article}, pdfauthor={Author}]{hyperref} 
\usepackage{amssymb}
\usepackage{amsmath}
\usepackage{subcaption}
\usepackage{xcolor}
\usepackage{caption}
\usepackage{soul}
\usepackage{microtype}
\usepackage{graphicx,lipsum}
\usepackage{upgreek}

\begin{document}

\title{Kinetic view on dynamic plasticity of crystalline solids}

\author{Elijah Borodin}
    \email[Correspondence email address: ]{elijah.borodin@manchester.ac.uk}
    \affiliation{School of Engineering, The University of Manchester, Manchester M13 9PL, UK}
\author{Afonso D. M. Barroso}
    \affiliation{School of Engineering, The University of Manchester, Manchester M13 9PL, UK}
\author{Andrey P. Jivkov}
    \affiliation{School of Engineering, The University of Manchester, Manchester M13 9PL, UK}

\date{\today}

\begin{abstract}
Microstructural changes in solids, driven by energy flows, do not develop in a static continuous space, such as the space considered in conventional plasticity models. The applied forces create an evolving internal energy landscape, which is constrained by crystallography but has characteristic spatial and temporal scales that form dynamically. To describe this view, we replace a common model for the evolution of dislocation substructure in metals with evolution of microscopic slips in a combinatorial structure referred to as a polytopal cell complex (PCC). The microscopic slips are associated with pairs of 2-cells (faces) and 1-cells (edges) of the PCC and are driven by minimisation of a properly defined Lagrangian. The approach provides a comprehensive statistical and thermodynamic description of plastic flow development. It allows the investigation of energy levels associated with different slip systems and reveals the microscopic mechanisms that result in the phenomenon of strain rate sensitivity.
\end{abstract}

\keywords{discrete dislocation plasticity, slip systems, polytopal cell complexes, dislocation kinetics, characteristic relaxation time, plastic flow thermodynamics}

\maketitle

\noindent Intense energy flows into materials generate combinatorial and topological changes in their internal structures at all length scales. 'Combinatorial' are the changes in the numbers of distinct structural components, e.g., an increasing number of phases or multiplication of dislocations as a result of plastic deformation. 'Topological' are the changes in the spatial arrangement and connectivity of structural components, such as networks of interphase boundaries, dislocation slips or microcracks. Combinatorial changes necessarily produce topological changes, but the inverse is not guaranteed. These changes redistribute, dissipate or store the injected energy producing the macroscopically observed stress-strain behaviour of a material. 
The relation between the combinatorial and topological changes and the resultant development of multiscale bulk plastic flow is far from being fully understood. A comprehensive thermodynamic description of plasticity requires a consideration of all kinetic and dynamic processes driving the internal microstructural changes.

The existing plasticity theories conventionally employ semi-empirical models that provide an averaged behaviour matching the macroscopic deformation curves. However, the importance of crystallographic features for metal plasticity was recognised relatively early and introduced in several constitutive models. The common Molinari approach \cite{Molinari1987, MolinariToth1994} uses an orientation factor:
\begin{equation} 
    r^{\beta}_{ik} = \frac{1}{2} \left( \nu_i \ell_k + \ell_i \nu_k \right)
        \label{eq: orientation_factor}
\end{equation} 
for each crystallographic slip system $\beta \in \mathbb{N}$, where $\vec{\nu}$ is the unit vector normal to the slip system, and $\vec{\ell}$ is the unit vector in the slip direction. The relation between the plastic strain rate $\dot{\varepsilon}^{\mathrm{p}}_{ik}$ and the applied \textit{deviatoric} stress $\sigma_{kl}$ is then given by \cite{Molinari1987,MolinariToth1994}
\begin{equation} 
    \dot{\varepsilon}^{\mathrm{p}}_{kl}
    = \dot{\varepsilon}_{0}\cdot \sum_{\beta} r^{\beta}_{kl} \left( \frac{\left|r^{\beta}_{pq} \sigma_{pq} \right|}{\sigma_0^{\beta}}\right)^{\gamma+1}
        \label{eq: const_Molinary}
\end{equation} 
where $\dot{\varepsilon}_{0}$ is a reference constant strain rate, $\sigma_0^{\beta}$ is a reference stress associated with the slip system $\beta$, and $\gamma$ is a coefficient of stress amplitude sensitivity \cite{Gruzdkov1999,Borodin2018}. The latter is related to the strain rate sensitivity $m$ \cite{Suo2013} via $\gamma = \left(m^{-1}-1\right)$. The three parameters cannot be derived theoretically for a wide range of deformation conditions and are typically obtained by curve fitting to experimental data. In many cases, such as nanocrystalline materials or very high strain rates, materials become extremely sensitive to strain rate \cite{Rida2020}, which render the constitutive equation unsuitable.
Structural approaches \cite{Mayer2013, Mayer2015, Malygin1999} explicitly accounting for the dynamics of dislocation gliding and the kinetics of dislocation multiplication \cite{Krasnikov2011, Krasnikov2010MD} replace Eq. \eqref{eq: const_Molinary} with the Orowan equation \cite{Hirth_new2017, Mayer2013, Krasnikov2011}:
\begin{equation}
	\dot{\varepsilon}^{\mathrm{p}}_{ik} = -\frac{1}{2}\sum_{\beta} \left( b_i^{\beta}\nu_k^{\beta} + b_k^{\beta}\nu_i^{\beta} \right) \rho_D^{\beta} V_D^{\beta},
        \label{eq: Orowan}
\end{equation} 
where $b_i^{\beta}$ are the components of the averaged Burgers vector \cite{Hirth_new2017}, $\rho_D^{\beta}$ is the averaged dislocation density \cite{Hirth_new2017, Landau_Mech}, and $V_D^{\beta}$ is the averaged dislocation velocity in the slip system $\beta$. $V_D^{\beta}$ can be estimated analytically \cite{Mayer2013} or by molecular dynamics simulations \cite{Krasnikov2010MD} as a function of the externally applied stress. The structural approaches based on Eq. \eqref{eq: Orowan} naturally incorporate the effects of strain rate sensitivity and allow for  macroscopic modelling of plastic flow even under shock wave deformation conditions \cite{Mayer2015, Krasnikov2011}. However, the exact calculation of the evolution of $\rho_D^{\beta}$ and $V_D^{\beta}$ during the deformation process for each slip system is not a feasible task, and only the averaged $\rho_D$ and $V_D$ values have been used in practice \cite{Mayer2015, Krasnikov2011, Borodin2015, Borodin2018}. 
As a result, the existing continuous plasticity models, from one side, have failed to handle complex non-equilibrium states produced by high defect concentrations and strong defect-defect interactions \cite{Walgraef2002, Malygin1999}, and from another, the microstructural parameters describing dislocation physics are not convenient for mechanical and engineering applications. 

In continuum models, the static yield stress can be generalised to a strain-rate-dependent parameter of dynamic yield strength \cite{Meyers1994}.
One approach is by considering an integral criterion of plasticity based on the 'fading memory' concept and using a material parameter $\tau_D$, referred to as the characteristic relaxation time (CRT) \cite{Gruzdkov2002,Gruzdkov2009,Petrov2007,Gruzdkov1999,Selyutina2016}:
\begin{equation}
	\frac{1}{\tau_D} \int_{t-\tau_D}^t \left(\frac{\sigma_{eq}(\tau)}{\sigma_y} \right)^\gamma d\tau \leq 1. \label{eq: IntegralCriterion_general} 
\end{equation}
Here, $\sigma_{eq}(\tau)$ is the equivalent stress at time $\tau$, $\sigma_{y}$ is the static yield stress, and the equality defines the time $\tau_D = t^*$ at which the macroscopic plastic flow begins.
It has been shown \cite{Selyutina2018,Borodin2017} that the CRT is inversely proportional to the dislocation density $\rho_D$:
\begin{equation}
	\tau_D = \frac{8 B_p}{3\mu b^2 \rho_D},
    \label{eq: CRT_rhoD}
\end{equation}
where $\mu$ is the material's shear modulus, and $B_p\approx 0.5 \nu_D\rho b^2$ is a phonon drag coefficient for dislocation gliding \cite{Selyutina2016, Krasnikov2011, Mayer2013} with $\nu_D\sim 10^{13}$ s$^{-1}$ being close to the Debye frequency and $\rho$ being the material density. 
This approach does not consider any crystallographic and microstructural features, but a generalisation of Eq. \eqref{eq: IntegralCriterion_general} suggested by Gruzdkov \mbox{\cite{Gruzdkov2002, Gruzdkov2009}} is suitable for multi-energy-level processes. It employs a series of CRTs $\tau^{\beta}_D$ with corresponding critical barrier stresses $\sigma_{\beta}$:
\begin{equation}
	\sum_{\beta} \left(\frac{1}{\sigma_{\beta}}-\frac{1}{\sigma_{\beta+1}}\right)\cdot \frac{1}{\tau^{\beta}_D} \int_{t-\tau^{\beta}_D}^t \sigma^{\beta}_{eq}(\tau)\cdot \theta(\tau) d\tau \leq 1, 
 \label{eq: IntegralCriterion_series} 
\end{equation}
where $\theta(\tau)$ is the Heaviside step function. The CRTs $\tau_D^\beta$, can be computed from Eq. \eqref{eq: CRT_rhoD} by replacing $\rho_D$ with the slip system-specific $\rho_D^\beta$.

The barrier stresses $\sigma_{\beta}$ are commonly estimated in crystal plasticity theories as critical resolved shear stresses determined by the generalised Schmid's law \cite{VanHoutte2005LAMEL}. However, in a purely continuous formulation, there is no possibility to define the dynamically changing values of the CRTs $\tau_D^{\beta}$. The value of the total dislocation density $\rho_D = \sum_{\beta}\rho_D^{\beta}$ strongly depends on the extremely complex dynamics of defect-defect interactions \cite{Walgraef2002}, including self-organisation of dislocations \cite{Malygin1999}, and can vary up to several orders of magnitude during severe plastic deformation \cite{Estrin2013, Borodin2019} or in materials subjected to high-strain-rate loading conditions \cite{Armstrong2010, Mayer2015, Selyutina2016}.


Naǐmark \textit{et al}. \cite{Naimark1992,Naimark1998,Naimark2017} proposed handling the t structure complexity by empirical and statistical considerations of plasticity, assuming a self-similarity of the spatial distributions of plastic defects \cite{Barenblatt2003}, $W(\lambda d ) = \lambda^{-\eta} W(d)$, \textit{i.e.}, postulating that the probability distribution function of defects $W(s,\vec{\nu},\vec{\ell})$ is invariant of the length scale $d$ \cite{Naimark1992} and has the form 
\begin{equation}
	W = Z^{-1} \, \mathrm{exp}(-E/Q),
            \label{eq: distribution_function_W}
\end{equation} 
where $E$ is the total energy (Lagrangian) of the defects, $Q$ is a potential specifying the energy relief, and $Z$ is a partition function \cite{Landau_SatPhysV}. A key idea in our present consideration is to use the microscopic plastic slip (microslip) event as the unit of plastic deformation \cite{Naimark1992,Naimark1998} instead of the explicit consideration of dislocations. By doing so, a 'real' network of one-dimensional dislocation lines is replaced by a virtual \emph{network of two-dimensional microslips} created by dislocation gliding. In some cases the microslips are associated with stacking faults stretched between two partial dislocations and have the corresponding stacking fault energy \cite{Hirth_new2017}, but in the general case the microslip structure should be viewed as a virtual quasi-periodic ordering in space formed by dislocation traces. 

In the continuum formulation, slips on the mesoscale level can be described by a microslip tensor $s_{ik}$ \cite{Naimark1992}, which quantifies the plastic strain in the neighbourhood of a local interface (the slip plane) \cite{Naimark1998}
\begin{equation}
	s^{\beta}_{ik} = s\cdot r^{\beta}_{ik}, 
            \label{eq: microshears_tensor}
\end{equation} 
where $s$ is the shear intensity. 

The microslips in a given volume produce a microslip density tensor given by \cite{Naimark1998}:
\begin{equation}
	p_{ik} = n \langle s_{ik} \rangle = n \int {s_{ik} W(s,\vec{\nu},\vec{\ell}) \: ds \: d^3\nu \: d^3 \ell}, 
            \label{eq: density_tensor_general}
\end{equation} 
where $n$ is the scalar concentration of microslips and $\langle s_{ik} \rangle$ is the statistical averaging of $s^{\beta}_{ik}$ \cite{Naimark1992, Naimark1998}. By definition, $p_{ik}$ is equivalent to the plastic strain $\varepsilon^{p}_{ik}$, however it has a larger theoretical significance as a measure of the energy $U$ accumulated in the dislocation cores and their local elastic fields. It may be referred to as the macroscopic microslip density tensor.

For a comprehensive thermodynamic description, it is essential to define the statistical distribution function Eq. \eqref{eq: distribution_function_W}. The Lagrangian $E$ depends on the kinetics of microslips and can be expressed in the form \cite{Naimark1992,Naimark1998}
\begin{equation}
	E = E_0 - H_{ik}s_{ik} + \alpha s^2_{ik}, 
            \label{eq: Lagrangian_continuous}
\end{equation} 
where $E_0$ is a constant 'ground energy', $H_{ik} = \sigma_{ik} + \lambda p_{ik}$ is the effective field which is the sum of an externally imposed stress field $\sigma_{ik}$ and of a contribution from the macroscopic microslip density tensor $p_{ik}$ \cite{Naimark1998}; the constant $\alpha$ describes a self-energy associated with the creation of one microslip, addressing the non-equilibrium nature of the process under consideration \cite{Naimark1998}. Integration of Eq. \eqref{eq: density_tensor_general} with Eq. \eqref{eq: Lagrangian_continuous} gives a strongly non-linear complex behaviour of the microslip density tensor $p_{ik}$, dependent on the particular stress field and on the dimensionless scaling factor $\delta = \alpha/\lambda n$ \cite{Naimark1998,Naimark2017}.

This promising statistical approach faces the fundamental restrictions of the continuous description of materials, which do not allow for considering the essential effects of local, multi-scale and multi-dimensional defect-defect interactions governing plasticity on the micro- and meso-scale levels. In particular, to complete the thermodynamic description (Eqs.\eqref{eq: microshears_tensor} -- \eqref{eq: Lagrangian_continuous}) and find the distribution function in Eq.\eqref{eq: distribution_function_W}, a partition function $Z$ has to be determined, which cannot be done without explicit consideration of the defect network and can only be replaced by an empirical fitting function, as it was done in \cite{Naimark1998,Naimark2017} using the Landau formalism for second-order phase transitions \cite{Landau_SatPhysV}.

The present work revisits the continuum statistical approaches using  combinatorial tools from graph theory \cite{Bollobas1998} and algebraic topology \cite{Kozlov2008}. It accounts for the constraints imposed by a specific defect ordering and provides an informative description of plastic flow development by computationally addressing the mesoscale problem. The results presented later are relevant to FCC metals, where \{111\} is the only active slip plane creating $12$ slip systems \cite{Kittel2018}. However, the theoretical consideration can be extended to the cases of several active slip planes in BCC and HCP crystals using associated combinatorial spaces.

Similarly to the continuous formulation discussed in \cite{Naimark1992,Naimark1998,Naimark2017}, we replace the dislocation network with a \emph{discrete space ordering} created by the tracks of gliding dislocations. This ordering is determined by the dislocation patterns and the dislocation sliding. The aim is to supplement the model for dynamic plasticity with an \emph{additional combinatorial structure} representing explicitly the features that redistribute and accumulate energy during the plastic flow. This is done by associating the microslips with the 2-cells of a polytopal cell complex (PCC) $\mathcal{T}$ \cite{Kozlov2008} and the network of microslips with an evolving graph $G$ where nodes represent microslip events and edges connect neighouring microslips. In such case, the partition function $Z$ is calculated directly using combinatorial techniques such as Tutte polynomials $T(G,x,y)$ \cite{Bollobas1998}. However, the Metropolis algorithm \cite{Sethna2006book, Zhu2024} is computationally less demanding than finding the Tutte polynomials for large graphs and will be employed in the current study. Here, we emphasise the distinction between our a priori 'discrete' space and the 'discretised' spaces obtained by meshing of a continuous domain: modelling by discretisation is a top-down approach, while the discrete modelling is a bottom-up procedure. For analysis of discrete spaces, we use tools from discrete exterior calculus (DEC) \cite{Hirani2003,Yavari2012,Berbatov2022}, a mathematical framework analogous to the exterior calculus with differential forms \cite{Hirani2003}. This framework has been significantly developed in recent years, allowing for direct applications in the mechanics \cite{Boom2022,Dassios2018,Jivkov2012} and physics \cite{Berbatov2022,Jivkov2023,Seruga2020} of solids.

\begin{figure}[ht]
	\centering
	\includegraphics[width=0.98\linewidth]{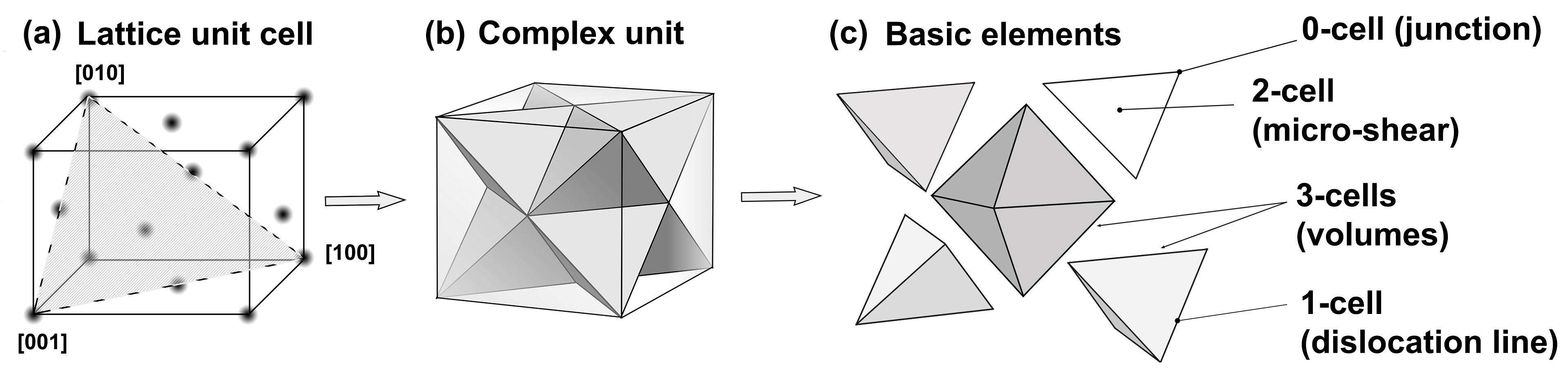}
	\caption{Sketch of a PCC construction associated with the 12 active slip systems in FCC metals.}
	\label{fig: FCC PCC PHYSICS}
\end{figure}

A discrete representation of the spatial arrangement formed by the lattice of \{111\} slip planes in a face-centred cubic (FCC) structure can be obtained with a PCC \cite{Jivkov2023, Borodin2024, Boom2022, Berbatov2023, BorodinJivkov} created from an appropriate tessellation $\mathcal{T}$ of a 3-dimensional continuum volume, as is shown in Fig. \ref{fig: FCC PCC PHYSICS}. Starting from the geometry of the FCC unit cell, Fig. \ref{fig: FCC PCC PHYSICS} (a), all 12 slip systems can be represented by considering eight tetrahedra and one octahedron in a closely packed arrangement, as shown in Fig. \ref{fig: FCC PCC PHYSICS} (b). The cell complex has four types of 'cells' of different topological dimensions, highlighted in Fig. \ref{fig: FCC PCC PHYSICS} (c) (see Appendix A for more details). The cells of the constructed structure represent volumes that may contain many crystallographic unit cells, but the structure has the same crystallographic restrictions of the material's internal order.

%

A PCC $\mathcal{T}$ \cite{Kozlov2008, Jivkov2012, Boom2022} constructed this way is a discrete space where physical processes may occur. Its topology and, once embedded in a metric space, geometry, impose specific restrictions on the plasticity process.  
The current deformation state of the system subjected to dynamical loading is encoded in a state vector $X_i$ whose components are zeros and ones indicating those microslip events that have (or have not) occurred. As previously stated, a graph $G$ (not a cell complex) can be constructed by considering the microslip events in $\mathcal{T}$ as nodes and connecting with edges those that are adjacent in $\mathcal{T}$ \cite{Bollobas1998, Durrett2010}. There is thus a three-way correspondence between each stage of the plasticity process, the state vectors $X_i$, and the evolution of the graph $G$. Each state vector is characterised by a particular discrete Lagrangian $E_p$ equivalent to the continuous energy parameter $E$ given by Eq. \eqref{eq: Lagrangian_continuous}.

To achieve a combinatorial description of the framework laid out in Eqs. \eqref{eq: distribution_function_W} -- \eqref{eq: Lagrangian_continuous}, we rewrite the microslip tensor $s_{ik}$ as a vector-valued 2-cochain, i.e. a function assigning a vector to a 2-cell (see Fig. \ref{fig: FCC PCC PHYSICS} (c)). Suppose the PCC is embedded in a metric space and that a microslip event is occurring on a particular 2-cell with area $A$ and local volume $V$; we define the local volume of a 2-cell as the sum of the volumes of the two 3-cells incident on that 2-cell. Because $s_{ik}$ is a measure of the plastic deformation induced by slip at 2-cell, we define the shear intensity $s=|\vec{b}|A/V$.
Furthermore, we replace the 2-cell normal $\vec{\nu}$, which is extrinsic to $\mathcal{T}$, by the oriented area $Ac^2$ which \textit{is} the 2-cell, where $c^2$ is the unit cochain that assigns a value of 1 to the 2-cell in question. Then, one can obtain a discrete version of Eq. \eqref{eq: microshears_tensor} by de-symmetrising the tensor $s_{ik}$ and writing
\begin{equation}
    S^2 = s\vec{\ell}\otimes A c^2 = \left( \frac{|\vec{b}|A^2}{V}\ \vec{\ell} \right) c^2,
    \label{eq: slip_cochain}
\end{equation}
where the superscript $2$ in $S$ and $c$ denotes the dimension of the cochain (see Appendix A), not a power. Note that the use of both a 2-cell $c^2$ and a 1-cell on its boundary which determines $\vec{l}$ means that the basic elements in our consideration are not only 2-cells themselves, but every pairs composed of one 2-cell and one 1-cell on its boundary. It is these pairs that are assigned 1s and 0s in the state vectors $X_i$ and that constitute the nodes in the graph $G$.

The self-energy of microslips, computed as $\alpha s_{ik}:s_{ik}$ in the continuous model \cite{Naimark2017}, is similarly computed in the discrete model by an inner product of the 2-cochain $S^2$ with itself, multiplied by a parameter. The inner product of $p$-cochains (see Eq.\eqref{eq: cochain_inner_product} in Appendix A) carries with it a physical dimension of $L^{d-2p}$; therefore, the result of $\langle S^2, S^2 \rangle$ has a physical dimension of $L^3$.

Similarly, the Cauchy stress tensor $\sigma_{ik}$ has been shown to be best described as a covector-valued 2-form \cite{Kanso2007}. We obtain a discrete description of this tensor by first considering it to be a tensor-valued 0-cochain (tensor field) that assigns the same tensor to every node in the complex. From this we obtain a covector-valued 2-cochain, which we denote as $\sigma^2$, by applying the flat operator $\flat$ described in Ref. \cite{Boom2022}. This cochain assigns a force vector to each 2-cell. Then, the mechanical energy term in the continuum model, $\sigma_{ik}:s_{ik}$, becomes $\langle \sigma^2, S^2 \rangle$ in the discrete model.

Unlike $s_{ik}$ and $\sigma_{ik}$, which are tensor fields, the mean-field tensor $p_{ik}$ is a global tensor obtained from the statistical averaging of all microslip tensors $s_{ik}$ (see Eq. \eqref{eq: density_tensor_general}). This is translated into the discrete framework as follows. Suppose that $k-1$ microslip events have already occurred in the system, each defined by a microslip 2-cochain $S^2_i$ on 2-cells $c_{2,i}$, $i\in\{1,...,k-1\}$. The mean-field acting on the $k$-th microslip event occurring on 2-cell $c_{2,k}$ takes the vector value
\begin{equation}
    \vec{P} = \frac{1}{N} \sum_{i=1}^{k-1} S^2_i(c_{2,i}),
    \label{eq: discrete_meanfield}
\end{equation}
where $N$ is the total number of microslip events that can occur. This value $\vec{P}$ is then assigned to the 2-cell $c_{2,k}$ giving the vector-valued 2-cochain $P^2 = \vec{P} \: c^2_k$, and the inner product $\langle P^2,S^2 \rangle$ gives the discrete analogue to $p_{ik}:s_{ik}$. Like $\langle S^2,S^2 \rangle$, the inner product $\langle P^2,S^2 \rangle$ has a physical dimension of $L^3$. The above formulation of the mean-field is finite, in contrast to Eq. \eqref{eq: density_tensor_general} which makes use of an integral.
This has an implication on how to count microslip events. 
In the discrete formulation, the set of points where microslips can occur is the countably finite set of all 2-cells in the PCC; hence, the density of microslips can be replaced with their fraction $n=(k-1)/N$, where $k-1$ is the number of microslips that have occurred.

The discrete Lagrangian can be written using the new formulations
of the discrete measure of the plastic strain $S^2$ (\mbox{Eq. \eqref{eq: slip_cochain}}) and the mean-field $P^2$ (\mbox{Eq. \eqref{eq: discrete_meanfield})}:
\begin{equation}
    E_i = E_0 + \alpha  \langle S^2 , S^2 \rangle -
                \langle \sigma^2 , S^2 \rangle -
              \lambda \langle P^2 , S^2 \rangle ,
    \label{eq: Lagrangian_discrete} 
\end{equation}
where the parameters $\alpha$ and $\lambda$ in the discrete case have physical dimensions of stress. The constant energy term $E_0$ can be chosen arbitrarily as only the energy difference $\Delta E = E_i - E_0$ affects the probability of new microslips occurring.

Eqs. \eqref{eq: slip_cochain}--\eqref{eq: Lagrangian_discrete} provide a complete discrete formulation necessary for simulating plasticity processes developing in a discrete PCC space. However, none of the terms included in the discrete Lagrangian \mbox{Eq. \eqref{eq: Lagrangian_discrete}} can be calculated before a spatial metric is introduced in the form of a characteristic length scale $l_c$. A PCC, being a purely combinatorial construction, has no metric, so $S^2$ in Eq. \eqref{eq: slip_cochain} and $P^2$ in Eq. \eqref{eq: discrete_meanfield} remain undetermined without the explicit embedding of $\mathcal{T}$ into a physical metric space, which allows the determination of the 2-cells' areas. Defining the distance $l_c$ as the diameter of a circle inscribed in a 2-cell, we can determine a scale for the 2-cell area $A$ as
\begin{equation}
        A = \pi {(l_c)}^2/ 4.
        \label{eq: A_metric}
\end{equation}
%


The choice of the length scale must be based on physical considerations. We take the average distance between dislocations $l_c$ as a nominal length covered by a moving dislocation between two resting points \cite{Hirth_new2017}. 
In Refs. \cite{Meyers2000,GalindoNava2012} it has been shown that at moderate and large plastic deformations the average distance between dislocations corresponds to the average dislocation cell size \cite{Malygin1999}, which can be estimated as 
\begin{equation}
    d = \kappa \rho_D^{-1/2},
    \label{eq: size_map}
\end{equation}
where the coefficient $\kappa \approx 15$ \cite{GalindoNava2012,Borodin2019}. Eq. \eqref{eq: size_map} provides a natural scale, which, in turn, leads to estimation of the plastic strain $S^2$ and energy $E_i$ associated with each 2-cell in a PCC.
In Ref. \cite{Naimark2017} a dimensionless structural 'scaling' parameter $\delta = l_c /l_n$ has been used, where $l_n$ is a length associated with the dislocation core size. Taking into account that the average distance between dislocations $l_c\approx d$ (see Eq. \eqref{eq: size_map}), the average dislocation core size is about the size of the Burgers vector \cite{Meyers1994,Kittel2018} $l_n\approx b$, and considering further the estimations provided in Refs. \cite{Naimark1998,Naimark2017} giving $\delta = \alpha/ \lambda n$, allows one to obtain $l_c$ as a function of the concentration of microslips $n$ as
\begin{equation}
        l_c = \vartheta\frac{\alpha b}{ \lambda n}. 
    \label{eq: lc_general}
\end{equation}
For a maximal concentration of defects $n=1$, the dislocation density should also reach an experimental maximum $\rho_D^{\mathrm{max}}\sim 10^{17}$ m$^{-2}$, which, according to Eq. \eqref{eq: size_map}, gives a length scale $l_c \sim 10^{-8}$ m. With $\alpha \lesssim \lambda \lesssim 10\alpha$, this gives $\vartheta \sim 10^2$ -- $10^3$. On the other hand, a minimum fraction of $n \sim 10^{-5}$ for relatively fine meshes gives, for $\vartheta \sim 10^3$, the minimal corresponding dislocation density $\rho_D^{\mathrm{min}}\sim 10^6$ m$^{-2}$ and $l_c \sim 10^{-2}$ m.

A Python package titled 'Plastic Slip Simulator' \cite{PSS_code} has been written for the creation of the PCC described above. It allows the simulation by a stochastic process of the evolution of the microslip structure towards an equilibrium state for a specific stress $\sigma^2$. Specifically, an equilibrium state is found by employing the Metropolis-Hastings algorithm (see Appendix B for details) \cite{Sethna2006book,Zhu2024}. The algorithm arrives at a configuration with a (local) minimum energy corresponding to a microslip network with a particular state vector $X_i$, which in turn corresponds to a Lagrangian value given by Eq. \eqref{eq: Lagrangian_discrete}. Another Python package titled 'PCC Analyser' \cite{PCC_Analyser2023} allows one to visualise $\mathcal{T}$ and study the structure of microslips appearing on its 2-skeleton. Both codes are freely available on the GitHub repository MATERiA codes \cite{MATERiA_codes}.

\begin{figure}[ht]
	\centering
		\centering
		\includegraphics[width=0.4\linewidth]{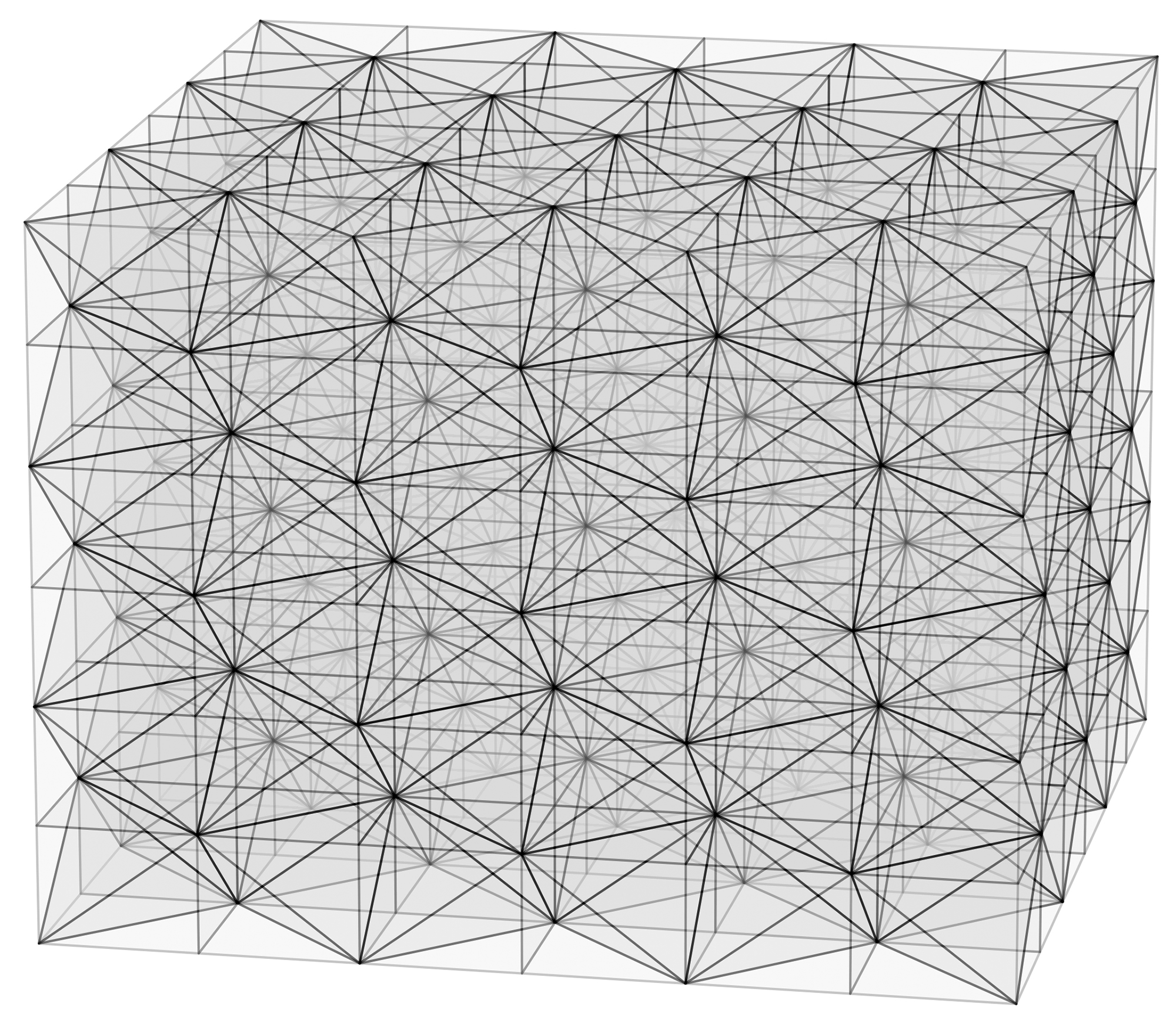} \hspace*{10pt} 
		\centering
		\includegraphics[width=0.4\linewidth]{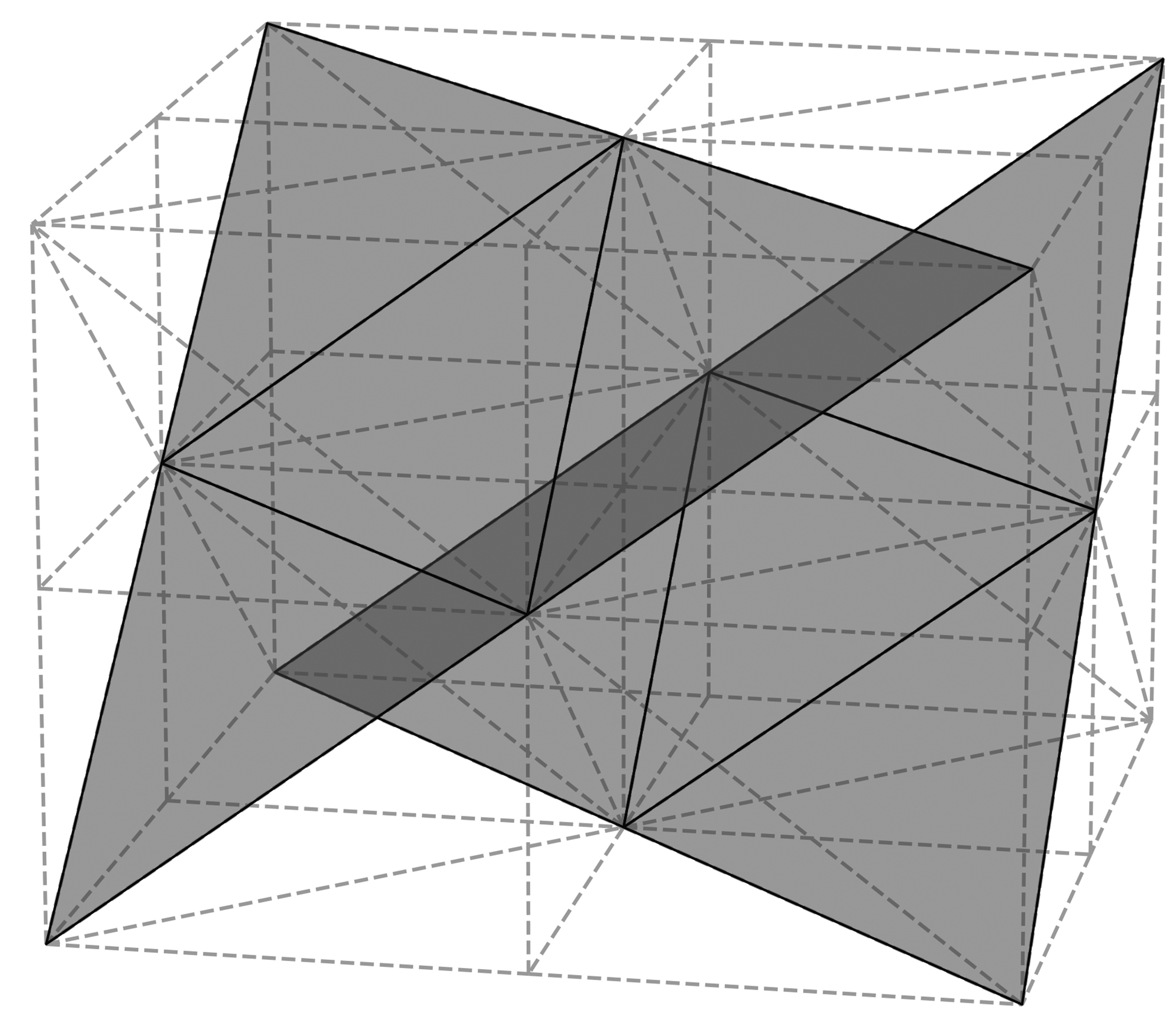}
	\caption{(left) The computational 3x3x3 PCC corresponding to the FCC crystals, and (right) the coloured $(1\bar{1}1)$ slip planes inside a unit cell -- each composed of four 2-cells (triangles).
 }
	\label{fig: FCC_cell_complex}
\end{figure}

\begin{figure*}[ht]
	\centering
    \begin{subfigure}{1\linewidth}
        \centering
    \begin{minipage}{0.31\textwidth}
        \centering%
		\includegraphics[width=0.97\textwidth]{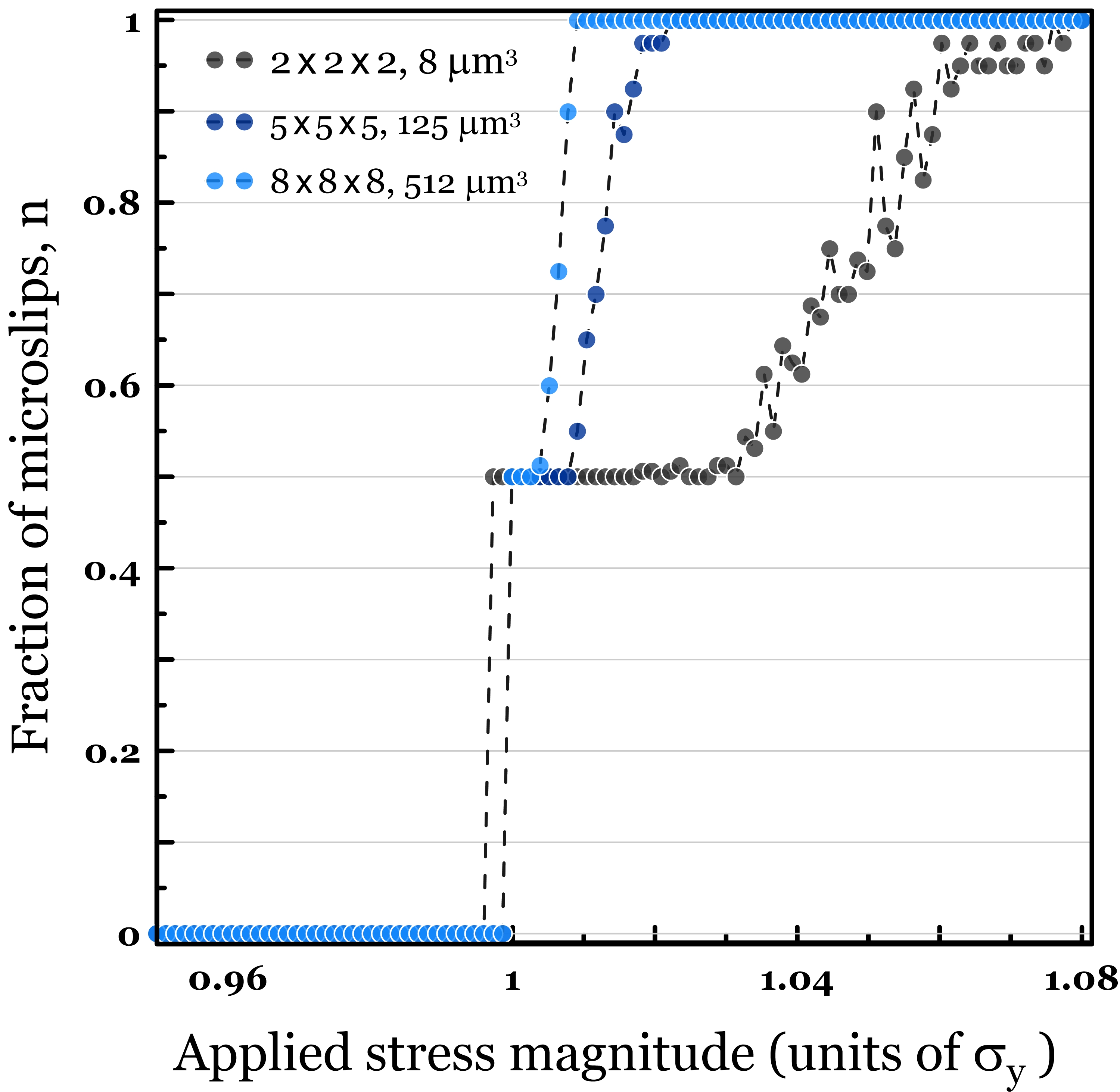} 
            \caption[short cap] {}
    \end{minipage}
    \begin{minipage}{0.44\textwidth}
            \centering%
		\includegraphics[width=0.97\textwidth]{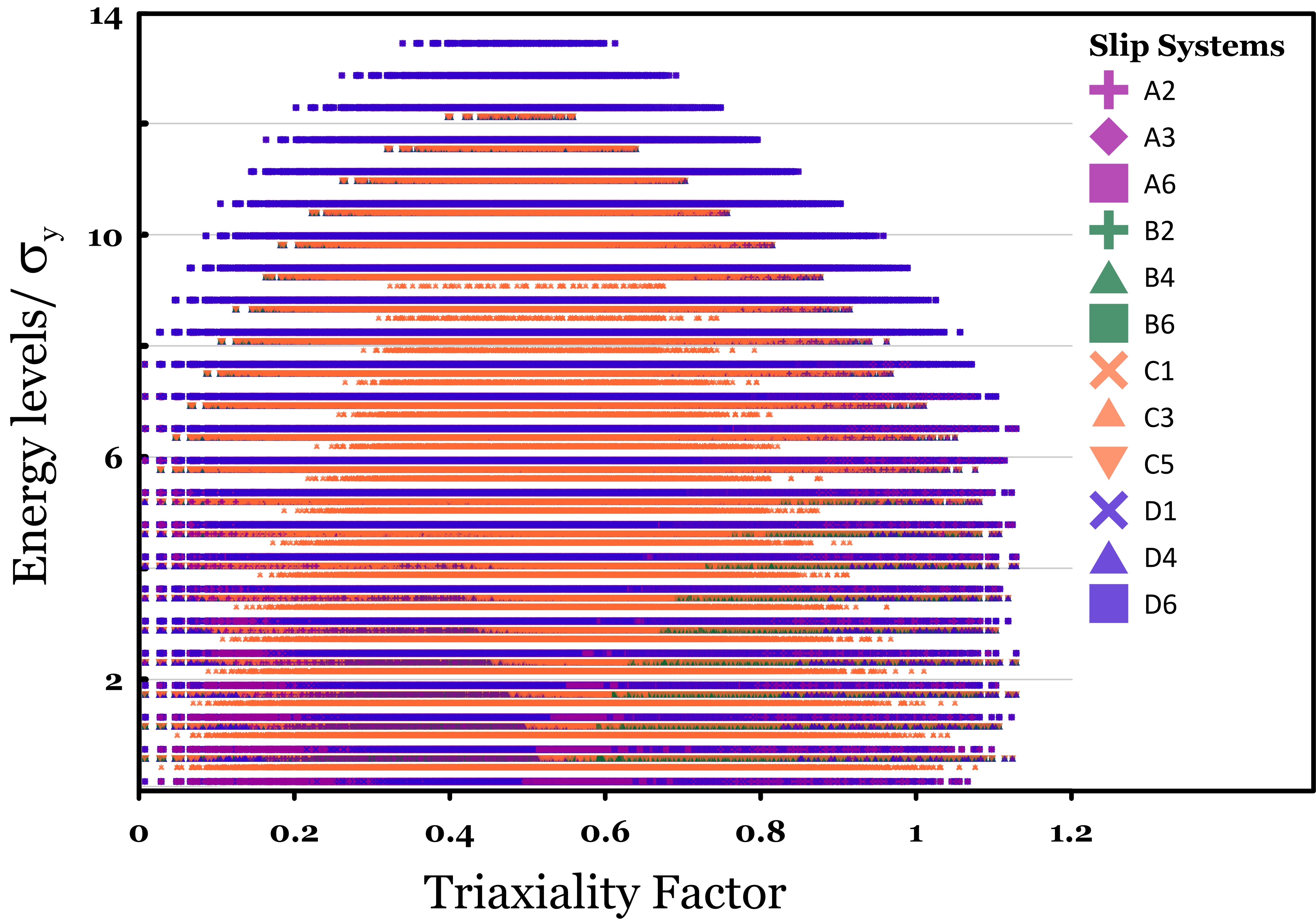}
            \caption[short cap] {}
    \end{minipage}
    \begin{minipage}{0.21\textwidth}
            \centering%
		\includegraphics[width=0.97\textwidth]{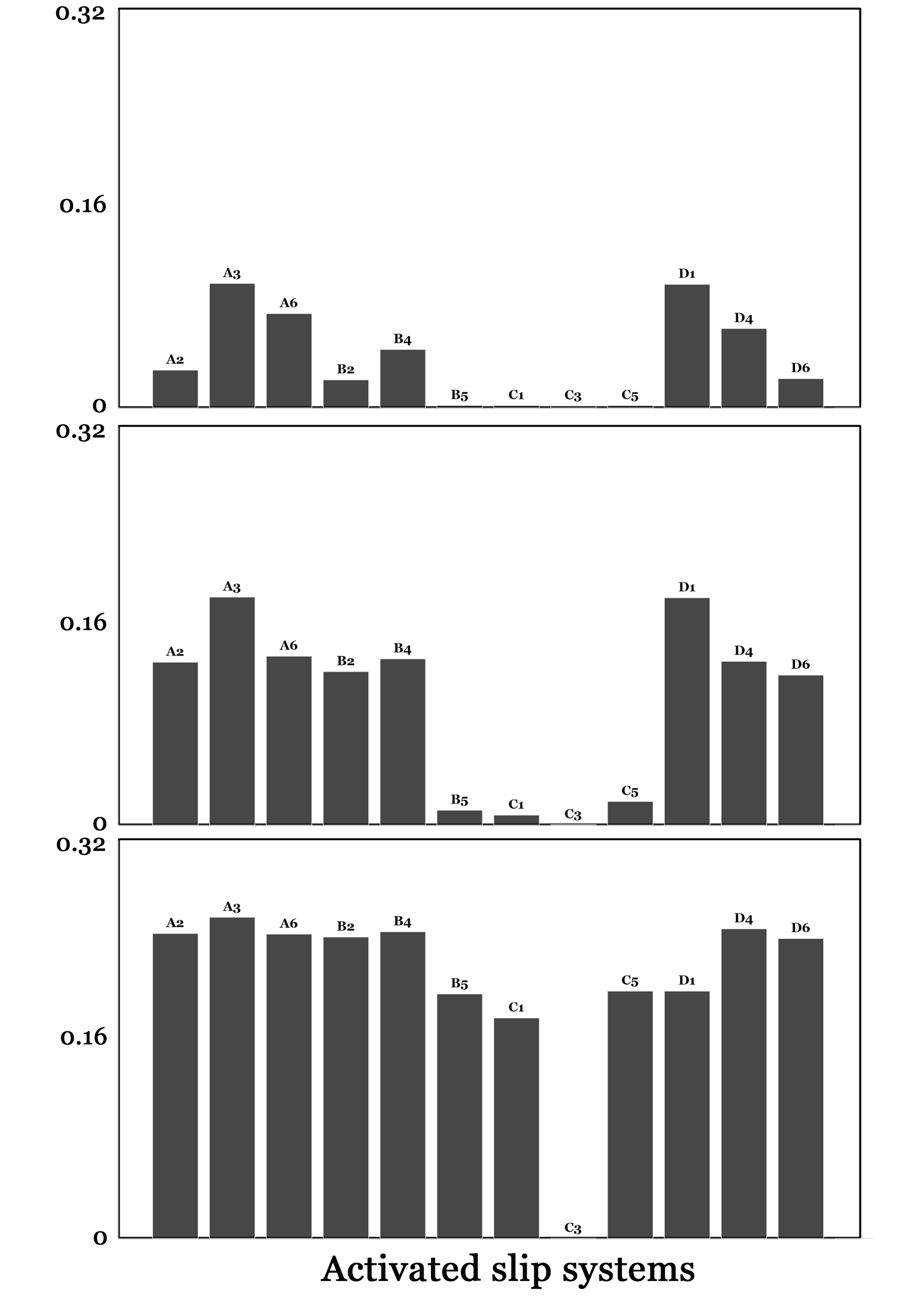}
            \caption[short cap] {}	
    \end{minipage}
\end{subfigure}

    \caption{(a) Dispersion of microslip fractions in PCCs of different sizes evolved from an initial state with zero microslips; (b) Density of the elastic energy levels as a function of stress triaxiality factor; and (c) Distributions of the inverse characteristic relaxation times $1/\tau^{\beta}_D$ in $\mathrm{ps}^{-1}$ (see Eq. \eqref{eq: IntegralCriterion_series}) over the activated slip planes during the plastic deformation process.}
	\label{fig: results}
\end{figure*}

Fig. \ref{fig: FCC_cell_complex} shows the simulation space with 27 unit cells which are supposed to represent a small part of the bulk material, revealing the topology of different slip systems.
Fig. \mbox{\ref{fig: results}}  shows an example of the simulation of monocrystalline pure copper under uniaxial tension with parameters specified in Appendix B.
The length scale $l_c$ proposed for the kinetic model implies that a \emph{representative plasticity volume} exists, which is the minimal volume where the steady-state plastic flow has a statistically smooth transition to a microslip saturation level.
It can be seen in Fig. \ref{fig: results}(a) that complexes smaller than $5 \times 5 \times 5$ computational unit cells have a statistically meaningful scattering in the fraction of microslips which suggests that the smaller complexes cannot provide enough 2-cells to maintain a statistically smooth development of the deformation process.  This is a consequence of a purely discrete formulation -- a bottom-up approach.
For a PCC inside a parallelepiped box with $\upsilon_{\xi}$, $\upsilon_{\upsilon}$ and $\upsilon_{\psi}$ unit cells in the three Cartesian directions, respectively, the total number of 2-cells corresponding to slip planes is given by
\begin{equation}
        N_2 = 32 \: \upsilon_{\xi} \upsilon_{\upsilon} \upsilon_{\psi}.
    \label{eq: 2cells_number}
\end{equation}
For a given metric scale $l_c$, the edge length $D$ of an appropriate cubical simulation space with $\upsilon = \upsilon_{\xi} = \upsilon_{\upsilon} = \upsilon_{\psi}$ can be computed from the geometric relationship
\begin{equation}
  D =
    \begin{cases}
      \frac{2 \sqrt{2} \upsilon l_c}{\sqrt{3}} & \text{if $\upsilon$ is even}\\
      \frac{2 \sqrt{2} \upsilon^2 l_c}{\sqrt{3\upsilon^2-2\upsilon+1}} & \text{if $\upsilon$ is odd}\\
    \end{cases}
\end{equation}
It is apparent from Eq. \mbox{\eqref{eq: lc_general}} that the metric scale $l_c$ and size of the associated simulation space can significantly change during the deformation process as the result of dislocation multiplication described by the microslip concentration $n$. This effect should be especially significant for large deformations \cite{Borodin2019} and high-strain-rate loading conditions \cite{Armstrong2010,Mayer2013,Krasnikov2011}. 
Using Eq. \eqref{eq: lc_general} and Eq. \eqref{eq: A_metric} for introducing a metric in the PCC, the cube ($\upsilon = \upsilon_{\xi} = \upsilon_{\upsilon} = \upsilon_{\psi}$) with linear size below $D_{crit}\approx 10 l_c$ becomes statistically not representative entailing the fact that the plastic flow development remains in non-equilibrium. For the saturation densities $\rho_D \sim 10^{14}$ m$^{-2}$ of severely deformed or dynamically loaded metals it gives $l_c \sim 1$ $\upmu$m and, correspondingly, $D_{crit}\approx 10$ $\mathrm{\upmu}$m.
More importantly, the complexes with a smaller subdivision demonstrate a pronounced step-wise curve (Fig.\mbox{\ref{fig: results}} (a)) indicating that the initial plasticity process generates a local internal mean-field $p_{ik}$ preventing its further development. This effect should become negligibly small for larger volumes of fine-grained materials.
Fig. \mbox{\ref{fig: results}}(b) shows how the number and values of the elastic energies associated with different slip systems (i.e. the term $\langle \sigma^2, S^2 \rangle$ in Eq. \eqref{eq: Lagrangian_discrete}) caused by an applied external stress magnitude $|\sigma| = 100$ MPa changes with the stress triaxiality factor $\eta$ \cite{Bao2004}:
\begin{equation}
    \eta = \frac{I_1}{3\sigma_{eq}}
        \label{eq: triaxiality}
\end{equation}
where $I_1$ denotes the first invariant of the Cauchy stress tensor \cite{Landau_Mech}. The graph demonstrates only the density distribution of 'allowed' energy levels, while the structure of energy gaps for each particular $\eta$ value can vary significantly. Slipping becomes possible on any slip system possessing an elastic energy above the self-energy barrier corresponding to $\alpha \langle S^2,S^2 \rangle$ in Eq. \eqref{eq: Lagrangian_discrete}. However, because of the probabilistic nature of the real deformation process -- expressed by using the Metropolis-Hastings algorithm in our computation model -- the self-energy is not a strict barrier and microslips have a non-zero probability to appear on 2-cells with energies slightly below it.

The calculations show (not presented in the figure) that at moderate stress values, most of the $\eta$ values correspond to 5-7 active slip systems with the limits of $2$ and $12$ active systems for a small number of values. It is interesting to note that even small changes in triaxiality factor can dramatically change the number and character of active slip systems, which can manifest itself in serrated stress curves like the Portevin–Le Chatelier (PLC) effect. At moderate stresses, the gaps between energy levels play a critical role due to the probabilistic nature of the process. 


The development of (discrete) plastic deformation can be represented as a sequence of the state vectors $X_i$, corresponding to the arrangement and fraction of the microslips in a PCC. At every evolution step, only a single 2-cell--1-cell pair in the PCC not yet included in the graph $G$ (meaning it is assigned a value of $0$ in $X_i$) becomes part of it by being assigned a value of $1$. However, such a structural evolution occurs entirely in the combinatorial space and has no direct correspondence to real time. An additional function is needed as a map of this combinatorial process into time $t$.  In the works of Naimark \textit{et al.} \cite{Naimark2017} such mapping was provided using the empirical Ginzburg-Landau theory of second order phase transitions \cite{Landau_SatPhysV}:
\begin{equation}
    \dot{n} = -\Gamma \frac{\partial F}{\partial n}\approx -\Gamma \left(\mathcal{A} \dot n - \mathcal{B}\dot n^3 + \mathcal{C}\dot n^5 - \mathcal{D}\sigma_{eq} n + \nabla (\mathcal{E} \nabla n) \right), 
    \label{eq: LandauEq}
\end{equation}
where $F$ is the free energy of the system and $\Gamma$, $\mathcal{A, B, C, D, E}$ are empirical parameters. Eq. \eqref{eq: LandauEq} maps space evolution of the defect-related structures with the concentration of microslips $n$ to the physical time; however, the large number of process-dependent fitting coefficients offers this few advantages and is unavoidable for a purely continuous formulation.
The discrete approach proposed in this work, in combination with the recently developed mechanical theory of CRTs (Eq. \eqref{eq: CRT_rhoD}) \cite{Selyutina2018} and the integral criterion of plasticity (Eq. \eqref{eq: IntegralCriterion_general}) \cite{Gruzdkov2009, Petrov2007, Gruzdkov1999}, allows for explicitly introducing a time scale. The CRT is associated with the time required for the development of a plasticity process in a representative volume (Fig. \ref{fig: results}(a)) and strongly depends on the kinetics of dislocations or the concentration $n$ of carriers of plastic slip \cite{Selyutina2018}. It can be stated that the time scale is being formed dynamically on micro- and mesoscale levels depending on the energy dissipation rate during the deformation process. 
If the average dislocation density in Eq. \eqref{eq: size_map} can be rewritten with  Eq. \eqref{eq: lc_general} as $\rho_D^{\beta} = \left(\kappa\lambda n_{\beta} /\vartheta \alpha b \right)^2$, then Eq. \eqref{eq: CRT_rhoD} for the CRT \cite{Selyutina2018} reads as 
\begin{equation}
	\tau^{\beta}_D = \frac{\Theta}{n^2_{\beta}},
    \label{eq: CRT_D}
\end{equation}
where the material-dependent coefficient $\Theta = 8\alpha^2 \vartheta^2 B_p/ 3\mu \kappa^2 \lambda^2$. For the considered set of parameters for copper alloys, $\Theta\sim 10^{-12}$ s for $\vartheta \sim 10^3$. The gradual increase of the microslip fraction $n$ (Eq. \eqref{eq: density_tensor_general}) during the deformation process decreases the characteristic relaxation times $\tau_D$, making the dynamic effect less pronounced \cite{Selyutina2016}.
Fig. \ref{fig: results}(c) shows the changes in the number of $\tau^{\beta}_D$ parameters and their values during the evolution of the system of microslips (top-down) towards a stationary plastic flow. Such transition stages are especially significant for extremely fast dynamic processes like ultrashort laser pulses or particle beam irradiation \cite{Mayer2015}, where the initial plastic flow instabilities can affect the whole deformation process. The estimated values of $\tau^{\beta}_D$ in the range of nanoseconds agree well with the values obtained in Refs. \cite{Selyutina2016, Selyutina2018, Selyutina2022} by fitting experimental data for bulk metallic samples. The proposed kinetic model reflects the spatio-temporal richness of the processes of plastic flow development.

Authors acknowledge the financial support from EPSRC, UK via grants EP/V022687/1 (PRISB) and EP/N026136/1 (GEMS). \hfill \break

\bibliographystyle{apsrev4-1}
\bibliography{kinetic_bibliography}

\appendix
\section{Appendix A: Elements of DEC} \label{app: DEC}

As a first introduction, a 3-complex $\mathcal{M}$ can be seen as a generalisation of a planar graph (itself a 1-complex) \cite{Grady&Polimeni2010, Kozlov2008}. Each $p$-dimensional object in an $n$-dimensional complex ($n \geq p \geq 0$) is called a $p$-cell. We will restrict ourselves to 3-dimensional complexes, which are composed of nodes (0-cells, points), edges (1-cells, lines), faces (2-cells, polygons) and volumes (3-cells, polyhedra). In a purely mathematical sense, a $p$-cell is homeomorphic to a closed $p$-ball \cite{Singh2019}. 
The connectivity of the elements in $\mathcal{M}$ is given by incidence matrices, $\partial_{p}$, referred also as the boundary operators, which relate $p$-cells to the $(p-1)$-cells on their boundaries for $p\in\{1, 2, 3\}$.
A $p$-cell has to be composed of a minimum of $p+1$ nodes --- a complex where the top-dimensional cells are exclusively composed of $p+1$ nodes is called simplicial. In 3 dimensions, this translates to the complex having only tetrahedra and triangles for 3- and 2-cells, respectively. 
Each $p$-cell in a cell complex is assigned an intrinsic orientation, which, in a simplicial complex, can be defined by an ordered listing of its constituent nodes \cite{Wilson2007}.
A $p$-cell with an \textit{intrinsic orientation} induces an \textit{induced orientation} on the $(p-1)$-cells on its boundary \cite{Hirani2003, Berbatov2022}. 
If a cell complex is embedded in a metric space, then it is possible to assign sizes to each $p$-cell, which are called measures; the measure of the $p$-cell $c_p$ is represented here as $\mu(c_p)$ after Berbatov \textit{et al.} \cite{Berbatov2022}.
A linear combination of $p$-cells is called a $p$-chain and represented with lowercase indices like $c_p$, and a function on $p$-chains is called a $p$-cochain and represented with uppercase indices like $c^p$. Using lower and upper indices to distinguish chains and cochains is reminiscent of their counterparts: vectors and forms. Like forms, cochains are anti-symmetric on reversal of a cell's intrinsic orientation. With this in mind, let us look again at the definition of the microslip 2-cochain in Eq. \eqref{eq: slip_cochain}.
Axial vectors are a construction of classical vector calculus which are better described by bivectors (or 2-blades) in the more general framework of geometric algebra, and which in turn find a discrete representation in 2-chains \cite{Hestenes}. Furthermore, unit axial vectors are intensive quantities, but we are interested in working with extensive quantities \cite{Jivkov2023}. Therefore, we replace the unit axial vector $\vec{\nu}$ in Eq. \eqref{eq: orientation_factor} by a 2-cochain $Ac^2$ that selects the 2-cell where slip occurs and multiplies by its area. Then, the tensor $s_{ik}$ is unsymmetrised and written in the discrete as a vector-valued 2-cochain $sA\vec{\ell}\otimes c^2$.
We take 3-cells to be representations of volume elements and assume that two neighbouring 3-cells slide over each other in the fashion of plastic slip, with the 2-cell shared by their boundaries as the interface. 

To multiply cochains, we use the inner product derived by Berbatov \textit{et al.} \cite{Berbatov2022}, which, for two general $p$-cochains $w^p$ and $q^p$ in a 3-complex, is given by
\begin{equation}
    \langle w^p , q^p \rangle =
    \sum_{c_p} \frac{w^p(c_p) \cdot q^p(c_p)}{4(p+1)\mu(c_p)^2}
    \sum_{n_0 \preccurlyeq c_p} \kappa(n_0)
    \sum_{v_3 \succcurlyeq n_0} \mu(v_3),
    \label{eq: cochain_inner_product}
\end{equation}
where $w^p(c_p)$ is the evaluation of the $p$-cochain $w^p$ on the unit $p$-chain $c_p$ (the same way one evaluates a function at a point), and $\kappa(n_0)$ is the weight assigned to the node $n_0$ \cite{Berbatov2022}. The notation $v_3 \succcurlyeq n_0$ is used to indicate that the node $n_0$ is on the boundary of the 3-cell $v_3$. Berbatov \textit{et al.} provide the weights $\kappa(n_0)$: since our cell complex has the shape of a parallelepiped, the weight of $n_0$ is 8 if it lies on a corner of the complex, 4 if it lies on an external edge of the complex, 2 if it lies on an external face, and 1 if it lies in the interior of the complex. This inner product is only non-zero if the two $p$-cochains take values on at least one and the same $p$-cell.

\section{Appendix B: Metropolis Algorithm for the Discrete Framework} \label{app: Appendix_Metropolis}

Simulations were conducted under uniaxial tension for pure copper, with $\sigma_{zz} \neq 0$ and all other components of the stress tensor equal to zero. The temperature was kept at 293 K and the Burgers vector was taken to be $0.2556$nm \cite{Simon1992}.
The material was assumed to yield at 100 MPa, but the character of the behaviour of the solution is universal. The mean-field coupling strength was set to $\lambda=2.8\alpha$.

\begin{enumerate}
	\item Start with any initial configuration;
	\item Randomly choose a face-edge pair -- the face will define $\nu^2$ and the unit vector in the direction of the edge will define $\vec{\ell}$;
	\item Attempt to flip the state of that pair, \textit{i.e.}, if this face-edge pair has not been slipped, then construct the slip 2-cochain $S^2$ according to Eq. \eqref{eq: slip_cochain}, and if it has been slipped then remove the corresponding slip 2-cochain from the system. In either case, make the necessary changes to $P^2$ according to Eq. \eqref{eq: discrete_meanfield};
	\item Compute the energy change $\Delta E$ in the system according to Eq. \eqref{eq: Lagrangian_discrete}, using $-S^2$ in the case of a slip event being undone:
	\subitem If $\Delta E < 0$, accept the change and add the 2-cell to the graph $\mathcal{X}_2$ if the slip is new, or remove it from the graph $\mathcal{X}_2$ if the slip event is being undone;
	\subitem If $\Delta E \geq 0$, produce a random number $p\in [0,1]$ and compute the probability
    $$\Pi=\exp\left({-\frac{\Delta E}{k_BT}}\right),$$
    where $k_B$ is Boltzmann's constant and $T$ is the temperature of the system.
    \item If $\Pi>p$, accept the proposed change and commit it to the system; if $\Pi<p$, return the slip state of the face-edge pair to its previous value and do not change $P^2$;
	\item Repeat from step 1 until the maximum number of iterations is reached.
\end{enumerate}

\end{document}